\documentclass{eptcs}
\providecommand{\event}{Refine 2011}
\usepackage{oz2e}

\title{Perspicuity and Granularity in Refinement}
\author{Eerke Boiten
\institute{
School of Computing, University of Kent, Canterbury, Kent, CT2 7NF, UK.}
\email{ E.A.Boiten@kent.ac.uk}}
\def\titlerunning{Perspicuity and Granularity in Refinement}
\def\authorrunning{E.A. Boiten}
\newcommand{\zdefs}{==}
\begin{document}
\maketitle
\begin{abstract}
This paper reconsiders refinements which introduce actions on the
concrete level which were not present at the abstract level. It draws a
distinction between concrete actions which are ``perspicuous" at the
abstract level, and changes of granularity of actions between different
levels of abstraction.

The main contribution of this paper is in exploring the relation between
these different methods of ``action refinement", and the basic refinement
relation that is used. In particular, it shows how the ``refining skip"
method is incompatible with failures-based refinement relations, and consequently 
 some decisions in designing Event-B refinement are entangled.
\end{abstract}

\section{Introduction}
This paper discusses how different ways of introducing ``extra" actions in refinement 
(such as weak refinement, action refinement, stuttering steps) relate to the underlying
refinement relations used (e.g.\ trace refinement, failures refinement). In particular, we aim to
show how the choices in those two dimensions are interdependent. The paper is not intended
to be polemic (``my formalism/refinement relation is better than yours") nor is it really meant to
be a first introduction to the topic. Where it appears to state the obvious, this is in an attempt to ensure
that commonalities, differences, and design decisions in refinement relations are exhibited
in an unambiguous and uncontroversial way.

Before describing the issues in detail, we consider an example. 
The example is presented in Z, but the notation used is not essential to
what follows in this paper. In general, most of what is described in this
paper could be expressed in ASM \cite{Schellhorn05}, (Event-)B \cite{Abrial10}, Z \cite{Woodcock96},  binary relations \cite{Derrick01},
UTP \cite{UTP98} or many other state-based
formalisms; for the moment
we make no assumptions about what refinement relation is ``in force".

This example is
due to Carroll Morgan, who presented it during an enlightening conversation at the 
2009 Dagstuhl seminar 
``Refinement Based Methods for the Construction of Dependable Systems".
The abstract specification is essentially a priority queue, stored
as a bag, so taking out an element involves selecting the minimum of 
the bag. Obvious specifications of functions $min$ on bags
and (later) $sorted$
on sequences are omitted. The schema $AS$ describes system states,
$AInit$ initial states, and the schemas $Ain$ and $Aout$ the operations
of adding and removing an element. The precondition $b \neq \emptybag$ is 
included explicitly in $Aout$, in recognition of it having to be an explicit guard
in alternative notations such as Event-B.
\begin{sidebyside}
\begin{schema}{AS} b: \bag \nat\end{schema}

\strut\\

\begin{schema}{Ain} \Delta AS \\ x?: \nat \ST b' = b \buni \lbag x? \rbag
\end{schema}
\nextside
\begin{schema}{AInit} AS' \ST b' = \lbag\,\rbag \end{schema}
\begin{schema}{Aout} \Delta AS \\ x!: \nat \ST
b \neq \emptybag \\
b = b' \buni \lbag x! \rbag\\
x! = min(b)
\end{schema}
\end{sidebyside}
The concrete specification uses a sequence to represent the queue.
Removing an element is only possible when the sequence is non-empty and sorted, in
which case the element to be removed is at the head of the sequence.
The schema $Sort$ describes the sorting of the sequence.
The schema $Cycle$ is mostly a red herring\footnote{One might use it to represent the non-determinism in a
distributed implementation where individual clients have no control over
the access pointer in a cyclical list, \ldots maybe.} and not part of Morgan's original example.
\begin{sidebyside}
\begin{schema}{CS} s: \seq \nat\end{schema}

\strut\\[-2ex]

\begin{schema}{Cin} \Delta CS \\ x?: \nat \ST s' = s \cat \lseq x? \rseq
\end{schema}

\strut

\begin{schema}{Sort}\Delta CS \ST
\items s = \items s' \\ sorted(s') 
\end{schema}
\nextside
\begin{schema}{CInit} CS' \ST s' = \emptyseq\end{schema}
\begin{schema}{Cout} \Delta CS \\ x!: \nat \ST
s \neq \emptyseq \\
sorted(s) \\ s = \lseq x! \rseq \cat s'
\end{schema}
\begin{schema}{Cycle}\Delta CS \ST
s = \emptyseq \wedge s' = \emptyseq\  \vee\\
s' = (\tail s) \cat \lseq \head s \rseq
\end{schema}
\end{sidebyside}

This paper discusses the many ways in which one may consider the concrete
specification to refine the abstract one, possibly after a slight modification, or
possibly not at all, depending on the notions
of refinement and action refinement employed. Before we move on to that level of
complication, consider the composed schema 
$SortOut
\zdefs 
Sort 
\fcmp 
Cout$, 
whose
meaning is given by
\begin{schema}{SortOut}
\Delta CS \\ x!: \nat \ST
s \neq \emptyseq \\
\exists s'': \seq \nat \dot \items s = \items s''
\wedge sorted(s'')
\wedge s'' = \lseq x! \rseq \cat s'
\end{schema}

Then, uncontroversially, in most sensible refinement relations, the
operation $Aout$ is refined by $SortOut$ (or more precisely: the data type
$(AS,AInit,Ain,Aout)$ is refined by $(CS,CInit,Cin,SortOut)$)
under the retrieve relation
$b= \items s$. In fact, this is normally an equivalence: refinement also holds in the
reverse direction\footnote{A refinement linking $Ain$ to $Cin \fcmp Sort$ instead
is equally possible but would require strengthening the concrete state
invariant to sorted sequences; $Cin \fcmp Sort$ then simplifies to
the insert operation of insertion sort.}.

The rest of this paper is structured as follows. In Section \ref{sec:reft} we describe
different basic refinement notions. Then in Section \ref{sec:add} we discuss the various methods in which ``extra" operations
may appear in refinement steps. In Section \ref{sec:action} we compare how these methods can be used to model the
decomposition of actions into smaller grained ones, and how this impacts on the various basic refinement notions. Finally, Section
\ref{sec:conc} presents some conclusions.

\section{Basic Notions of Refinement}\label{sec:reft}
We have given detailed fully formal descriptions and comparisons of the different basic notions of
refinement for state-based and concurrent systems in many previous papers, e.g. \cite{BDS07,Derrick01,3023}. 
Rather than repeating this and thereby fixing a formalism or even introducing a new one, we
remain informal here, using various formalisms and their refinement notions as illustrations.

In basic data refinement, systems (or machines or abstract data types) are compared which have
identical alphabets (or sets of labels of operations (or actions or events)). Apart from conditions on initial
and possibly final states, and other details which depend on what observations can be made of these systems,
operations are compared in pairs of an abstract and a concrete operation, 
with refinement conditions being some subset of the following properties:
\begin{description}
\item[(1) Consistency] The effect of the concrete operation is one that is allowed by the abstract operation.
\item[(2) Enabledness] When operations can be invoked in the abstract state, they can be invoked in the concrete state as well.
\item[(3) Restricted consistency] In states where the abstract operation is enabled, the effect of the concrete operation is one that is
allowed by the abstract operation.
\end{description}
Property (1) or its weaker variant (3) represents the essence of refinement: that a client would be unable to observe conclusively
that they are using the concrete rather than the abstract system. Property (2) ensures that the client is indeed able to perform
the same ``experiments" on both systems.
Property (1) obviously  implies (3), and also a converse of (2): where concrete operations are enabled (leading to an ``effect"), their abstract counterparts should be
enabled, too (in order to allow comparison of effects). The properties leave out detail about what an effect is, are purposefully vague on ``can be invoked" in (2) to allow a variety of interpretations, 
and leave any linking between abstract and concrete states implicit. They are also somewhat
biased towards downward simulation.  A few examples should make all this clearer. The refinement relations described below will be refered to in later sections.

Traditional  (downward simulation) {\em Z refinement} \cite{Woodcock96,Derrick01} is characterised by 
properties (2) and (3), with ``can be invoked" in a state computed
as individual operations' preconditions, i.e.\ whether their defining predicates can be satisfied for some after-state.
Condition (2) is called ``applicability" and typically formulated as
\[ \pre AOp \wedge R \imp \pre COp \]
where $\pre AOp \zdefs \exists AS' \dot AOp$ denotes the computed precondition. 
Condition (3) is called ``correctness", and typically formulated as
\[ \pre AOp \wedge R \wedge COp \imp \exists AS' \dot R' \wedge AOp \]
We have sometimes called this refinement relation the ``contract" model of refinement as it
constrains the implementation only within the original precondition.

{\em Trace refinement} is characterised by (1) only, only requiring that anything that {\em does} happen in the concrete
specification is consistent with the abstract one. As such, it represents preservation of safety properties only, ``nothing bad happens".
No concrete operations being enabled at all, for example,  is an acceptable trace refinement.

Basic {\em Event-B refinement} (called simple refinement in \cite[Ch.\ 14]{Abrial10}) is characterised by (1), with (optionally) a 
weak alternative to (2): if the concrete state deadlocks (i.e.
no events are enabled), then so should the abstract state. Enabledness of events is given by explicitly specified guards, with a
``feasibility" proof obligation ensuring that they are at least as strong as any computed precondition. Abrial \cite[p.\ 429]{Abrial10} states
that condition (2) could be imposed, but ``this happens to be sometimes too strong". (We will return to this.)

{\em Failures-based variants} of refinement are characterised by (1) and (2), where (2) considers individual operations for ``blocking
Z refinement" and singleton failures refinement, or sets of concurrently enabled operations for failures refinement
as in CSP. We refer to \cite{BDS07,Reeves08,3023} for detailed discussion of these refinement relations and the finer distinctions
between them, which are not relevant in the current paper.

Note that a refinement relation characterised by property (3) without property (2) is nonsensical as it is not transitive: preconditions or
guards can be strengthened (lack of (2)) and then weakened (by (3)), but the composition of such steps does not respect (3).

\section{Adding Operations in Refinement}\label{sec:add}
The basic refinement rules described above deal only with the situation where the abstract and concrete specifications have
the same alphabet of operations. There are many ways in which one could allow a refined specification to have ``extra" operations -- we discuss a number of them. First, we mention alphabet extension and alphabet translation 
\cite[Ch.\ 14]{Derrick01} for completeness. Then, we get to the core of this paper: stuttering steps, the
introduction of internal operations, and action refinement, and how these sometimes get conflated.

\subsection{Alphabet Extension and Translation}
The simplest way of allowing new operations in refinement is {\em alphabet extension}: to just accept them without any further constraints. If we make the intuitive
step of identifying a non-existent operation with one that is never enabled, alphabet extension should be perfectly acceptable in
traditional Z refinement: it means we allow implementors to provide functionality that we had not asked for. In a process algebra context
alphabet extension is typically not allowed, and indeed that would make sense in our intuitive view: it would go against 
refinement property (1), by having no matching abstract behaviour for some concrete behaviour.

In {\em alphabet translation}, a single abstract operation is implemented by multiple concrete ones, which requires
an explicit mapping, recording for every concrete operation which abstract operation it represents, and thus which operation's behaviour it needs
to correspond with. (If this mapping is not required to be total, alphabet extension is subsumed.) A typical example for this would be
an abstract two-dimensional grid specification with a ``move" operation, which is refined into ``moveNorth", ``moveEast", etc. Alphabet translation is allowed in Event-B, where it is called ``splitting" an abstract event.

The semantic property established in alphabet translation is: every concrete trace (with its corresponding observations) is
consistent with an abstract trace that relates to it by the given mapping (applied elementwise) with its corresponding
observations.

\subsection{Perspicuous Operations}
State-based systems potentially change state when operations are executed. When no operation is invoked, the state does not normally change. Some formalisms take this into account by including explicitly so-called stuttering steps in their semantics: steps where the state does not change between two observations, due to no event having taken place. In the light of that, it is intuitively obvious to accept the introduction
of additional concrete events as refinements of the identity operation (a.k.a.\ $skip$) on the abstract state. We will call these 
{\em perspicuous} concrete events, to be distinguished from ``internal events" (see below) which incur additional assumptions and
requirements. In particular, in subsequent refinement steps, perspicuous operations do {\em not} have a different status from operations
that were present earlier.

Abrial \cite{Abrial10} presents a similar motivation for the introducion of new events in Event-B, analogous to how this is done
in action systems \cite{Back93}, and refers to it as ``observing our discrete system in the refinement with
a finer grain than in the abstraction". Event-B is explicit about the introduction of such events as being refinements of {\em modelling}:
introducing not just aspects of a solution, but more detail of the model. Indeed, where refinement is viewed as only moving from
a complete description of a problem to its solution, the introduction of perspicuous operations which achieve nothing in the abstract
world can hardly be useful by itself\footnote{This is {\em not} intended to be a controversial statement or implicit
criticism on Event-B: the crux is in the phrase {\em by itself}, and this should become clearer later when we compare the different
ways of encoding action refinement.}. Both action systems and Event-B include a relative deadlock freedom condition with this kind
of refinement: the new system should deadlock (i.e., terminate, in the action systems view) no more often than the old one.
The semantic relation established by this kind of generalised refinement is: for every concrete trace with its corresponding observations,
an abstract trace constructed by crossing out all perspicuous actions is consistent with it.

In the running example, under most refinement relations and with the obvious retrieve relation $\items s = b$ both concrete operations $Sort$ and $Cycle$ are candidate 
perspicuous operations, as they satisfy $\items s = \items s'$ and thus relate identical abstract states. They are both applicable in
every concrete state and thus are refinements of an abstract $skip$ even when property (2) is imposed.

For perspicuous operations, the notion of {\em divergence} comes into the picture. A collection of perspicuous operations is divergent if infinitely often in succession, from some state, one of its members can be invoked. In a trace-based view, where perspicuous operations could be inserted at arbitrary points between ``normal" operations, non-divergence is necessary to ensure that a finite trace cannot get extended into an infinite one by that
process. This is how Abrial \cite{Abrial10} explains it\footnote{His use of the term ``reachable" is a bit unfortunate, though -- this tends to be an existential property (some path is finite) rather than the required universal (all paths are finite) property required.}. With
additional assumptions, such as that a system might perform perspicuous operations independently, divergence becomes
a practical as well as a theoretical problem. Butler \cite{Butler09}
explains the non-divergence requirement in Event-B by saying ``The new events introduced in a refinement step can be viewed as
hidden events not visible to the environment of a system and are thus outside
the control of the environment" which would suggest these are not just perspicuous events, but even {\em internal} events as we will
discuss next. In action systems \cite{Back93}, which are viewed as a main inspiration for Event-B, all actions could be considered to be internal (even if the variables they modify are not), which conforms more with Abrial's explanation than with Butler's\footnote{Note
however that Abrial \cite{Abrial10} does recognise (on page 414) a different class of operation that ``{\em is not part of the protocol: it corresponds to a ``daemon" acting \ldots}".}.
A typical method of proving non-divergence is by establishing a variant (well-founded, strictly decreasing function)
on newly introduced (collections of) perspicuous operations \cite{Butler97a,Derrick98a,Abrial10}. If refinement is based on property
(1) rather than property (3), i.e., an action cannot gain behaviour in refinement, then non-divergence is preserved by subsequent
refinement steps.

In the example, both perspicuous operations are divergent. This is obvious from the fact that they are enabled in {\em every} concrete
state. $Sort$ allows an infinite sequence of invocations of which only the first
does not necessarily correspond to a concrete $skip$. For formalisms that use infinite traces and allow stuttering steps, such as TLA, this may not be a problem. Removing divergence on each of the operations can be done using several possible small modifications.
The divergence problem for $Sort$ could be fixed by including a guard $\neg sorted(s)$, but this makes it a refinement of 
$skip$ only if property (2) is not imposed and guards can be strengthened. 
Another way would be to add a flag that ensures $Sort$ is invoked exactly once
after every occurrence of $Cin$ or $Cycle$ (possibly also preventing the next $Cin$ until after sorting). A counter could be used to remove divergence in $Cycle$, with each of the other operations
(excluding $Sort$) setting the counter to fix the maximal number of occurrences of $Cycle$ to follow it, and $Cycle$ decrementing
it at every step until it is $0$. None of those modifications would retain the property that $Sort$ or $Cycle$ refines $skip$ if
the prevalent refinement relation respects (2).

\subsection{Internal Operations}
An internal operation is a perspicuous operation with a special status: it is assumed to be invisible to the environment, and under internal
control of the system only. In process algebras, internal operations naturally occur in a number of ways. In CSP \cite{Hoare85a}
they arise from channels being hidden, for
example encapsulating an internal communication channel when considering a system of communicating subsystems. They may also be
used, for example in LOTOS \cite{lotos1}, to encode internal choice when only external choice is available as a basic operator. Butler first considered the introduction of internal events in B refinement
\cite{Butler97a}, and based on this approach we introduced ``weak refinement" for Z \cite{Derrick98a,Derrick99a},
which was analysed and compared to ASM refinement in detail by Schellhorn \cite{Schellhorn05}. 

The requirements imposed in this context are inspired by how process algebras deal with internal actions, for example in defining
``weak" bisimulation: where standard refinement conditions refer to a single action, their ``weak" equivalents consider the same action
possibly prefixed and postfixed by occurrences of internal actions. Thus, the refinement consistency property, e.g.,  will state that for every concrete
action, with internal concrete behaviour before and after, its effect is consistent with the abstract action, possibly also pre- and postfixed
with (abstract) internal behaviour. E.g. in \cite{Derrick98a} the restricted consistency (correctness) condition for weak refinement in Z
(downward simulation) is phrased as
\[
\pre (Int_A \fcmp AOp) \wedge R \wedge (Int_C \fcmp COp \fcmp Int_C) \imp \exists AS' \dot R' \wedge (Int_A \fcmp AOp \fcmp Int_A)
\]
where $Int_C$ is arbitrary internal behaviour in the concrete state, i.e. the transitive reflexive closure of the union of internal
operations, and similar for $Int_A$. Taking this process algebra inspired approach has a few consequences:
\begin{itemize}
\item internal actions have a special status which goes beyond the refinement step where they are introduced. They can not only be introduced this way, but must also be taken into consideration or can even be removed in subsequent refinement steps.
\item there is an assumption that if internal actions are necessary for progress, they will ``eventually" happen, so external operations
are viewed as ``enabled" if their before-state is reachable through internal behaviour; in timed
process algebras in particular, internal actions are often taken as ``urgent" meaning they happen as soon as they are enabled.
\item there need not be independent refinement conditions for internal operations: all internal behaviour is viewed in the context of
its composition with external behaviour. Thus, internal operations need {\em not} be refinements of $skip$. Of course, all
internal operations being perspicuous, with external operations corresponding as normal, is {\em one} way of satisfying 
the refinement conditions like the one above, but it is not the only way. In fact, in some refinement relations, it may not be a viable way,
 see below.
\end{itemize}
The approaches for B and Z mentioned above only included {\em prevention} of divergence in weak refinement steps. A more general approach,
also consistent with the process algebraic view, is to {\em  preserve} or {\em reduce} any divergence that was already present in the abstract
specification.  This is worked out in detail in \cite{BDS07}, and the impact of differing notions of ``livelock" or divergence is discussed
in \cite{2838}. The semantic relation established in this case is roughly that for every concrete trace, an abstract trace exists that is consistent with it, with both traces' subsequences of {\em external} actions being identical\footnote{In fact it is a somewhat more subtle matching: non-determinism included in a single operation on one abstraction level may be represented through a different choice of sequence
of internal actions on the other level, so it is really a relation between sets of abstract  vs.\ concrete traces with the same external
subsequence.}.

\subsection{Action Refinement}
Alphabet translation described above allows for arbitrary matchings of an occurence of an abstract action with the occurrence
of a single concrete action. The most explicit
way of changing the granularity of actions is to allow for matchings between {\em sequences} of abstract and concrete actions.
This has been called ``action refinement" \cite{Aceto92} or ``non-atomic refinement" \cite{Derrick99a}. 
In its most\footnote{Avoiding for now the generalisation to $m$-to-$n$ diagrams with $m \neq 1$.} general form, action refinement corresponds to ASM $1$-to-$n$
diagrams with $n$ possibly greater than $1$ \cite{Schellhorn05}, generalising the normal commuting
simulation diagram to one where the concrete effect is achieved in $n$
steps, without requiring a relation between abstract and {\em intermediate}
concrete states. In this view, all concrete operations resulting from
the decomposition are of the same status, with only their order 
having an impact on refinement conditions.
This is also the view we took
in definining non-atomic refinement for Z \cite{Derrick99a}, work which was continued by Derrick and Wehrheim \cite{1628}. 
This kind of action refinement
is even possible without changing the state space involved. It requires
an explicit matching between abstract actions and concrete action sequences,
which also extends to traces. The semantic relation aimed for is that concrete traces are consistent with abstract traces under this
extended matching relation. The concrete and the abstract models end up having different interfaces with this approach -- this may be
exactly what is required, though. For example, \cite[Ch.\ 13]{Derrick01} has an example of a watch which in the abstract model has a
$ResetTime$ operation, which in the concrete model is represented by a series of executions of $ButtonA$ and $ButtonB$
operations.

Considering for simplicity now only the case that $n=2$, the refinement requirements are like the introduction of
sequential composition in refinement calculus \cite{Morgan94a}. Splitting an operation in two means finding an intermediate
state (predicate) such that the first ``half" lands in the intermediate state, and the second ``half" moves from the intermediate
to the original after-state. The problematic issue is what is or is not allowed to happen in the intermediate state.
In a concurrent context, this comes under the heading of ``interference" -- when the first ``half" of an operation
has been executed, should other operations be disabled (non-interference, as e.g.\ discussed for action systems in \cite{Back93}), or
should their execution cancel out the effect of this one? This is a well-known problematic area, discussed also in \cite{Derrick99a}, which
we will not focus on here, as it is orthogonal to the issues discussed: when an action is split with part of it
being perspicuous or internal, that also creates an intermediate state with the same potential interference problems.

\section{How to Reduce Granularity in Refinement}\label{sec:action}
From the discussion above, it should be clear that there are at least three semantic models for reducing the granularity of actions
in refinement:
\begin{itemize} 
\item by introducing perspicuous actions that take on some of the ``work" -- possibly requiring non-divergence;
\item by introducing internal actions to the same effect -- either using the limited refinement rules for perspicuous actions, or
by using the more general ``weak refinement" rules;
\item by giving explicit decompositions of actions in which all parts have the same status.
\end{itemize}
We limit ourselves for now to the case where we are decomposing an action into two actions, where the first part could be viewed
as ``prepatory work", and the second part as the ``real work" -- in other words, the situation in our example
of refining $Aout$ into $Sort$ and $Cout$, where we expect $Sort$ to be executed before $Aout$. However, in order to concentrate
on the general situation, let us consider refining $AWork$ into $Prepare$ and $CWork$.

For the methods of reducing granularity by refining $skip$, we aim for $Prepare$ to be perspicuous, and for $CWork$ to be a refinement
of $AWork$.  Now consider an abstract state in which the operation $AWork$ was applicable. If in every corresponding concrete state it would be possible to apply $CWork$, then we have a degenerate situation: we are introducing a new action $Prepare$ whose contribution is
unnecessary in all situations (i.e., it might as well be a {\em concrete} $skip$, too). Thus, in any relevant case of reducing granularity, $CWork$ can be applicable  in only a subset of the corresponding concrete states -- namely those where $Prepare$ has nothing (left) to do. Indeed,
because $Prepare$ is a refinement of an abstract $skip$, if its before-state is linked to a particular abstract state, then so should its
after-state. Again in order to ensure that $Prepare$ does something useful in some circumstances, there should be some abstract states
linked to the before-states of $Prepare$.

This is where the prevalent notion of refinement makes a difference. If condition (2) (``enabledness") is in force, we have made it impossible
for $CWork$ to be a refinement of $AWork$, because $CWork$ is only applicable in a strict subset of the corresponding concrete states.
This holds a fortiori for stronger versions of condition (2) such as failures refinement.

Thus, condition (2) excludes reduction of granularity by introducing perspicuous actions. It also excludes reduction of granularity by introducing internal actions using the ``perspicuous actions" conditions. However, the more general ``weak refinement" rules can be
used in combination with condition (2), as we have shown in \cite{BDS07} in a context with condition (1) in force, and in
\cite{Derrick99a} with condition (3) in force. This is explained by not being constrained to considering the concrete operation in
isolation, but rather only considering it in the context of possible internal concrete behaviour.

The other way in which condition (2) is problematic for the refinements of $ski$p is any requirements for perspicuous actions to be
non-divergent. If they are refinements of $skip$ respecting condition (2), then they are by definition applicable in all states and thus
always applicable ``again" and by definition divergent.

Returning to the example,  ignoring $Cycle$ for now, refinement reducing granularity is possible in several ways:
\begin{itemize}
\item by having $Sort$ perspicuous, and guarded by $\neg sorted(s)$ if it is also required to be non-divergent. This works for trace
refinement (just (1)), Event-B refinement, but not the other forms.
\item by having $Sort$ internal, provided it is guarded by $\neg sorted(s)$. This works according to the rules for Event-B, establishing
normal Event-B refinement. However,  it can also work for stronger refinement
relations respecting condition (2), but then the more general weak refinement rules need to be used to establish it. In particular,
it would mean that $Aout$ is compared for refinement with $Sort^* \fcmp Cout$.
\item for explicit action refinement of $Aout$ by $Sort$ followed by $Cout$, there is no requirement for $Sort$ to be guarded
(compare the watch example referred to above: as conceptionally the user presses $ButtonB$, there is no guard preventing
the user from doing that infinitely often), and refinement can be any kind, including relations respecting property (2) or even (3).
In fact, including a guard on $Sort$ would disallow the combined concrete output operation on states which are already sorted,
and thus be unacceptable if the refinement relation obeys property (2).
\end{itemize}

\section{Conclusion}\label{sec:conc}
The paradox that led to the discussion with Carroll Morgan referred to earlier was the following.
If the work of one abstract operation is split between two concrete ones, and one of the concrete operations makes no progress
that can be detected abstractly\footnote{Thus, some degree of data refinement is implied: a refinement of $\mathit{skip}$ on the {\em same}
state really cannot make any progress.}, why do we need this action at all? And if we do need it,
how can the other concrete operation, achieving some but not all of the
work of its abstract counterpart, be a refinement of the abstract one?
The answer is hopefully somewhat clarified above. It requires a notion of refinement that allows for guards to be strengthened.
The underlying issue may well have been known in ``folklore" but it is not presented in any published papers we are aware of.

Coming back to Event-B specifically, two of its design decisions are thus closely entangled:
\begin{itemize}
\item to have essentially a trace semantics with only global deadlock prevention;
\item to use stuttering step refinements for reducing granularity.
\end{itemize}
Both lead to relatively simple refinement obligations, which is attractive. In order for Event-B to strengthen refinement
to preserve stronger properties such as encoded in various refusal-based semantics, it would also have to give up its simple notion
of reduction of granularity. It could do this in at least two ways: either by going the way of ASM and having explicit recipes for
decomposing operations with their corresponding conditions, or by going the way of process algebra, and giving certain operations
explicit ``internal" status which they then would need to retain subsequently. In either case, the price of gaining semantic strength
is a considerable amount of complication of refinement conditions, which may be too big a price to pay, particularly for a formalism which now
has so much (automated) proof tool support available. Would that be what Abrial had in mind when he wrote that (condition (2)) ``happens
to be sometimes too strong"?

\subsection*{Postscript}
Finally, returning to the running example once more, a last word on the $Cycle$ operation. It makes no useful progress whatsoever,
but the constraints put upon this completely irrelevant operation in refinement in any ``stuttering steps" approach  (namely: taming its divergence), have been
no more and no less than on the supposedly enormously useful $Sort$ operation. Surely that is somewhat disappointing.

\subsection*{Acknowledgements} To Carroll Morgan for his explanations,
to  Michael Butler, John Derrick, Steve Dunne and Gerhard Schellhorn for useful discussions, and to the reviewers for
their comments.

\bibliographystyle{eptcs}
\bibliography{abz10}
\end{document}